# Discovery of a red backsplash galaxy candidate near M81

Kirsten J. Casey[⦿],[1,2]⋆ Johnny P. Greco,[1,2] Annika H. G. Peter[⦿][1,2,3,4] and A. Bianca Davis[1,2]

[1]*Physics Department, The Ohio State University, 181 W. Woodruff Ave., Columbus, OH 43210, USA*
[2]*Center for Cosmology and AstroParticle Physics, The Ohio State University, 181 W. Woodruff Ave., Columbus, OH 43210, USA*
[3]*Astronomy Department, The Ohio State University, 181 W. Woodruff Ave., Columbus, OH 43210, USA*
[4]*School of Natural Sciences, Institute for Advanced Study, 1 Einstein Drive, Princeton, NJ 08540, USA*

**ABSTRACT**
Understanding quenching mechanisms in low-mass galaxies is essential for understanding galaxy evolution overall. In particular, isolated galaxies are important tools to help disentangle the complex internal and external processes that impact star formation. Comparisons between quenched field and satellite galaxies in the low-mass regime offer a substantial opportunity for discovery, although very few quenched galaxies with masses below $M_\star \sim 10^9 M_\odot$ are known outside the virial radius, $R_{\rm vir}$, of any host halo. Importantly, simulations and observations suggest that an in-between population of backsplash galaxies also exists that may complement interpretations of environmental quenching. Backsplash galaxies – like field galaxies – reside outside the virial radius of a host halo, but their star formation can be deeply impacted by previous interactions with more massive systems. In this paper, we report the concurrent discovery of a low-mass ($M_\star \sim 10^7 M_\odot$) quenched galaxy approximately $1 R_{\rm vir}$ in projection from the M81 group. We use surface brightness fluctuations (SBF) to investigate the possibility that the new galaxy, dw0910+7326 (nicknamed Blobby), is a backsplash galaxy or a more distant field galaxy. The measured SBF distance of $3.21^{+0.15+0.41}_{-0.15-0.36}$ Mpc indicates that Blobby likely lies in the range $1.0 < R/R_{\rm vir} < 2.7$ outside the combined M81–M82 system. Given its distance and quiescence, Blobby is a good candidate for a backsplash galaxy and could provide hints about the formation and evolution of these interesting objects.

## 1 INTRODUCTION

Galaxy quiescence is a vital component to our understanding of galaxy evolution. The quenching of star formation is governed by a wide selection of complex processes, both internal and environmental. It is dependent on the removal, consumption, or fragmentation of a galaxy's cold gas reservoir. Internal mechanisms that affect galaxy quiescence include active galactic nuclei, supernovae, and stellar feedback, which heat up and sometimes eject a galaxy's cold gas (Croton et al. 2006; Hopkins, Quataert & Murray 2012; Christensen et al. 2016; Fitts et al. 2017; Cortese, Catinella & Smith 2021). These mechanisms are most often invoked for central galaxies because environmental effects dominate for low-mass satellite populations (Cortese et al. 2021). Relevant environmental mechanisms include interactions with the UV background, ram pressure stripping, tidal stripping, starvation, and preventive feedback (Gunn & Gott III 1972; Larson, Tinsley & Caldwell 1980; Boselli & Gavazzi 2006; Fillingham et al. 2015, 2018; Lu et al. 2017; Simpson et al. 2018; Cortese et al. 2021; Jahn et al. 2022; Wright et al. 2022). Although much of the literature is focused on high-mass galaxy quenching (Gabor et al. 2010; Wetzel et al. 2013; Peng, Maiolino & Cochrane 2015; Bluck et al. 2020; Li et al. 2020; Trussler et al. 2020; Guo et al. 2021; Oman et al. 2021; Trussler et al. 2021), for the purposes of this paper we will focus on dwarf galaxies, which we define as galaxies below $\sim 10^9 M_\odot$ in stellar mass. Much of the work on dwarf galaxy quiescence in recent years has been focused on disentangling the effects of these internal and environmental processes.

There are two main factors that determine whether any given galaxy is quenched or not: mass and environment. In the field, low-mass galaxies are preferentially star-forming and the fraction of star-forming galaxies increases with decreasing mass down to some limit imposed by reionization. Geha et al. (2012), for example, found that essentially no quenched field galaxies existed in their sample between stellar masses of about $10^{7-9} M_\odot$, while simulations by Jahn et al. (2022) and Samuel et al. (2022) suggest that this may hold true down to masses around $10^{6.5} M_\odot$. To date, only a few quiescent and isolated dwarf galaxies in this mass range have been discovered (Lavery & Mighell 1992; Whiting, Hau & Irwin 1999; Karachentsev et al. 2001; McConnachie et al. 2008; Makarov et al. 2012; Karachentsev, Kniazev & Sharina 2015; Makarova et al. 2017; Polzin et al. 2021; Sand et al. 2022). The trend with mass for field galaxies seems to result from the ongoing availability of gas falling on to isolated galaxy haloes and the inefficiency of internal quenching mechanisms to heat that gas (Munshi et al. 2013; Fitts et al. 2017). The exceptions to this trend are very low-mass field galaxies, often called reionization fossils. These galaxies have stellar masses below $\sim 10^5 M_\odot$ and their shallow potential wells are unable to maintain cold gas reservoirs after heating due to the UV background in the early Universe (Ricotti & Gnedin 2005; Gnedin & Kravtsov 2006; Sand et al. 2010, 2022; Jeon, Besla & Bromm 2017; Rey et al. 2020, 2022; Applebaum et al. 2021).

⋆ E-mail: casey.395@osu.edu

The story is quite different for low-mass galaxies in dense environments, where environmental processes dominate quiescence. In this regime, low-mass galaxies are much more likely to be quenched than their isolated counterparts, and the fraction of quenched satellites increases with decreasing mass (Mao et al. 2021; Greene et al. 2022). Environmental quenching mechanisms for satellite galaxies can be broken up into two main categories: processes that remove gas, and those that prevent the infall of new cold gas (Cortese et al. 2021). The former category includes ram pressure stripping and tidal stripping. Ram pressure stripping typically happens on short time-scales (Wheeler et al. 2014; Fillingham et al. 2015) and is caused by the pressure exerted on a satellite's interstellar medium as a result of its motion through the host's gaseous halo (Gunn & Gott III 1972; Simpson et al. 2018). Tidal stripping results from the gravitational interaction between a satellite and its host or even between satellites in a dense environment. Observational evidence of gas stripping is well established (Martínez-Delgado et al. 2001; Grebel, Gallagher III & Harbeck 2003; Mayer et al. 2007), and simulations suggest that this method of quenching dominates for satellites in the range $M_\star \sim 10^{6-8} M_\odot$ (Akins et al. 2021). Starvation, on the other hand, cuts off a satellite's access to its cold gas supply after infall into the host halo. This causes quenching on long time-scales (Larson et al. 1980; Garling et al. 2020) because the galaxy can continue to form stars with the remaining gas it has. The influence of starvation has been observed using spectroscopic data from the Sloan Digital Sky Survey (Peng et al. 2015; Trussler et al. 2020) and is demonstrated in hosts with halo masses as small as a few times $10^{11} M_\odot$, about the size of the Large Magellanic Cloud (Garling et al. 2020).

There is another population of low-mass galaxies important to the question of quiescence that remains relatively understudied, and that is the population of galaxies outside the virial radius of host haloes but close enough to be distinct from truly isolated field galaxies. In the simulations of Buck et al. (2019), for example, 'nearby' galaxies are defined as those between 1 and $2.5 R_{\rm vir}$ of the host, and field galaxies are those outside $2.5 R_{\rm vir}$. Importantly, the stellar-to-halo mass relation, gas content, and star formation histories (SFHs) of the simulated galaxies in this intermediate distance range more closely resemble satellite galaxies than field galaxies. Some of the difference between nearby galaxies and field galaxies could result from simplistic definitions of the halo boundary (More, Diemer & Kravtsov 2015; Diemer 2021), but at least some of the distinction seems to result from the presence of so-called backsplash galaxies (Benavides et al. 2021), which have entered the virial radius of a host galaxy in the past and now lie outside the virial radius of that host. Backsplash galaxies are distinct from truly isolated field galaxies and galaxies on first infall which have not at any point had host halo interactions even if they reside at similar distances from potential hosts.

The inclusion of backsplash galaxies outside the simple satellite-field dichotomy is important for understanding dwarf galaxy quenching. Their SFHs can be deeply impacted by interactions with the host and because they can end up quite far from the host halo they may confuse quenching analyses that only consider satellite and field populations. In cosmological hydrodynamical simulations by Benavides et al. (2021), for example, backsplash galaxies exist on average 2.1 $R_{\rm vir}$ and as much as 3.35 $R_{\rm vir}$ from the host, well into 'field' territory if only the distance is considered. In addition, the interaction that backsplash galaxies have with their hosts are enough to quench star formation (Lotz et al. 2019; Applebaum et al. 2021), though the interaction times these galaxies experience are generally smaller than those of satellites. Studies of SFHs and quenching time-scales between the two populations could provide hints about dominant quenching mechanisms in each scenario.

In terms of sheer numbers, backsplash populations are expected to be non-negligible. Simulations by Applebaum et al. (2021) have shown that for hosts similar to the Milky Way (MW), backsplash galaxies can make up a significant fraction (over 50 per cent in their case) of the galaxies in the range $1 < R/R_{\rm vir} < 2.5$. Similar results have been shown for larger groups and clusters, although these studies typically do not consider the dwarf galaxy regime (Rines et al. 2005; Pimbblet 2011; Wetzel et al. 2014; Borrow et al. 2023). Interestingly, there is some evidence from FIRE simulations that backsplash populations are not as abundant for low-mass hosts (Jahn et al. 2022). All of this highlights the importance of carefully considering the backsplash population of a given host when analysing low-mass galaxy quenching as a function of environment.

This paper reports the concurrent (Karachentsev & Kaisina 2022) discovery of a low-mass quenched galaxy with a projected distance that puts it in the outskirts of the M81 system (see Fig. 1 and the bottom-right panel of Fig. 2). We argue that the system is likely quenched based on its red $(B - R) \sim 1.2$ colour and because it has no discernible UV detection, which is often used as a tracer of recent star formation (see Section 4). Stellar population synthesis modelling suggests that this galaxy, dw0910+7326, has a stellar mass of the order of $10^7 M_\odot$. If the galaxy is isolated from the M81 system, then understanding its SFH is important for the sake of field galaxy evolution models. If it is close enough to M81 then it could be an example of a backsplash galaxy, a sparsely observed population, especially outside a Local Group (Teyssier, Johnston & Kuhlen 2012; Buck et al. 2019; Blaña et al. 2020; Applebaum et al. 2021) or cluster (Gill, Knebe & Gibson 2005; Lotz et al. 2019; Haggar et al. 2020; Benavides et al. 2021) context, and at such a low mass. The galaxy's name comes from Karachentsev & Kaisina (2022) indicating its position in the sky, but we will often refer to it by its nickname, Blobby.

The bare minimum that we need to determine whether Blobby is a backsplash galaxy or not is the distance. Surface brightness fluctuations (SBF) offer a promising technique to determine the distance to semiresolved galaxies where we have only photometric data. Stated simply, SBF takes advantage of the fact that galaxies appear smoother (have more stars per resolution element) when they are farther away (Fig. 2). We can therefore recover a distance by measuring the variance in pixel values across an image. SBF measurements break down where stars in a galaxy are not Poisson distributed, but SBF has been reliably applied in many galaxies from giant ellipticals (Tonry & Schneider 1988; Tonry, Ajhar & Luppino 1990; Jensen, Tonry & Luppino 1998; Blakeslee, Ajhar & Tonry 1999; Cantiello et al. 2018) to (more recently) low-mass galaxies (Jerjen, Freeman & Binggeli 2000; Jerjen, Binggeli & Barazza 2004; Mei et al. 2005; Cohen et al. 2018; van Dokkum et al. 2018; Carlsten et al. 2019, 2021a; Greco et al. 2021). Since the SBF method relies on semiresolved stellar populations, the distance range in which it can be reasonably used varies with the details of the observation and the stellar mass of the galaxy. For instance, Greco et al. (2021) estimate that for dwarf galaxies SBF can be used from a few Mpc out to approximately 25 Mpc with observatories like the upcoming Rubin Observatory. We refer the reader to Greco et al. (2021) for a more complete explanation of SBF in the low-luminosity regime.

The paper is organized as follows. Section 2 describes the observations and data reduction. Section 3 explains how we modelled the structure and SBF of the galaxy and how we validated the procedures used to do so. Section 4 describes the modelling and SBF distance measurement for Blobby itself. Finally, Section 5 discusses the implications of the measured distance, followed by concluding remarks in Section 6.

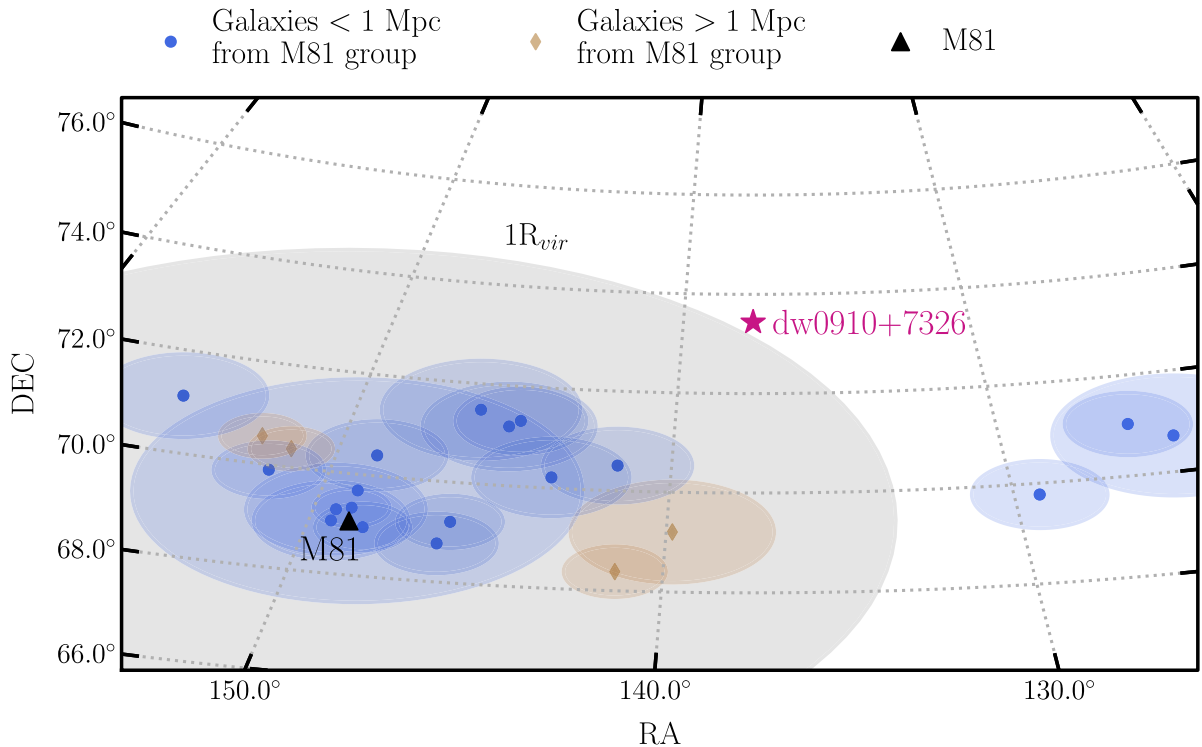

**Figure 1.** Field showing the relative positions of all galaxies in the Catalog of Local Volume Galaxies (Karachentsev & Kaisina [2019](#)) within a projected angle of 6 deg from dw0910+7326 (magenta star). Galaxies with distances in the range ∼3.0–4.3 Mpc in the M81 group or within approximately 1 Mpc of the M81 group are indicated with blue circles. There are four galaxies within the given projected radius but outside the M81 group, indicated with tan diamonds. These have distances in the range ∼7.5–9.0 Mpc from the Sun. M81 is indicated with a black triangle. The shaded regions around each point indicate the approximate virial radii for each galaxy. The large grey shaded region is the approximate virial radius of the combined M81–M82 system. For galaxies where the stellar mass is uncertain or there are disagreements from various sources, parameters that resulted in the largest virial radius were chosen in order to more conservatively estimate the relative proximity to dw0910+7326.

## 2 IMAGING AND DATA REDUCTION

The dwarf galaxy, dw0910+7326 (Blobby), was identified by chance in the DESI Legacy Imaging Surveys Data Release 9 Interactive Map[1] (Dey et al. [2019](#)). Our group was in the process of identifying galaxies for SBF analysis using the Large Binocular Camera (LBC; Ragazzoni et al. [2006](#); Speziali et al. [2008](#)) on the Large Binocular Telescope (LBT; Hill [2010](#)). The apparent low luminosity, isolation, and quiescence of this galaxy made it a particularly interesting candidate for follow-up observations. It also appeared semiresolved, indicating that it was likely in the Local Volume. Additional data and results for other low-mass galaxies will be presented in upcoming papers. Blobby's position with respect to other galaxies on the sky is shown in Fig. [1](#). Each nearby (background) galaxy is marked with a blue circle (tan diamond).

First, we investigated whether Blobby's projected distance to other galaxies meant that it was plausibly associated with any of them. For M81 (black triangle in Fig. [1](#)) we used the halo mass of the full M81–M82 system from Karachentsev & Kudrya ([2014](#)) ($\log M_{200} \sim 12.7$). Here, we define the halo mass by the 200 times critical density criterion. The halo mass was used to calculate an approximate virial radius, $R_{vir}$ (∼353 kpc for the M81–M82 system). Where galaxies from the Catalog of Local Volume Galaxies (Karachentsev & Kaisina [2019](#)) could be matched with the Atlas of Local Galaxies (Leroy et al. [2019](#)), the stellar mass from the latter was used. For the remaining galaxies we used NED[2] to find $g$- or $I$-band luminosities and $B - V$, $g - r$, or $V - I$ colours, which were then converted to $M/L$ ratios and finally stellar masses using Jester et al. ([2005](#)), Sharina et al. ([2008](#)), McGaugh & Schombert ([2014](#)), Willmer ([2018](#)), García-Benito et al. ([2019](#)), or Du & McGaugh ([2020](#)) relations, depending on the available data. For the few galaxies where this information was not available, we conservatively estimated $M/L_B$ of 10 (García-Benito et al. [2019](#)) to get the largest plausible virial radius. We then calculated a stellar mass using $B$-band luminosities derived from magnitudes in the Catalog of Local Volume Galaxies. Halo masses were calculated using a Moster abundance matching model (Moster, Naab & White [2013](#)). In cases where different authors disagreed on the properties of various galaxies, we used the literature values that resulted in the largest virial radii because the purpose of this endeavor was to determine possible galaxy interactions with Blobby. Fig. [1](#) makes it clear that even if Blobby is at the same distance as M81, and therefore the closest possible 3D distance to M81, it is still outside M81's virial radius and relatively isolated from all other known galaxies.

Follow-up observations of Blobby were conducted using the R-BESSEL and B-BESSEL filters of the LBC. These filters were chosen because we initially intended to use the $B - R$ SBF magnitude–colour relation from Jerjen et al. (2000) and Jerjen et al. (2004), and because SBF is particularly bright in *R*-band. The SBF method assumes that stars in a galaxy are Poisson distributed and therefore that the number of stars in a resolution element depends on the distance. The SBF signal, $\sigma^2_{\rm SBF}$, depends on the stellar luminosity function of the galaxy's stellar population. It has units of luminosity and is traditionally quantified as

$$\sigma^2_{\rm SBF} \equiv \bar{L} \equiv \frac{\sum_i n_i L_i^2}{\sum_i n_i L_i}, \qquad (1)$$

where the stellar population contains $n_i$ stars of luminosity $L_i$ (Tonry & Schneider 1988). The SBF magnitude, described in more detail in Section 3, is derived directly from $\sigma^2_{\rm SBF}$. It follows that more luminous stars, like those on the red giant branch and asymptotic giant branch, contribute more to the image variance, which makes red and near-IR filters particularly popular for SBF measurements (e.g. Blakeslee 2012; Greco et al. 2021).

---

[1] Legacy Surveys / D. Lang (Perimeter Institute) https://www.legacysurvey.org

[2] The NASA/IPAC Extragalactic Database (NED) is funded by the National Aeronautics and Space Administration and operated by the California Institute of Technology. https://ned.ipac.caltech.edu

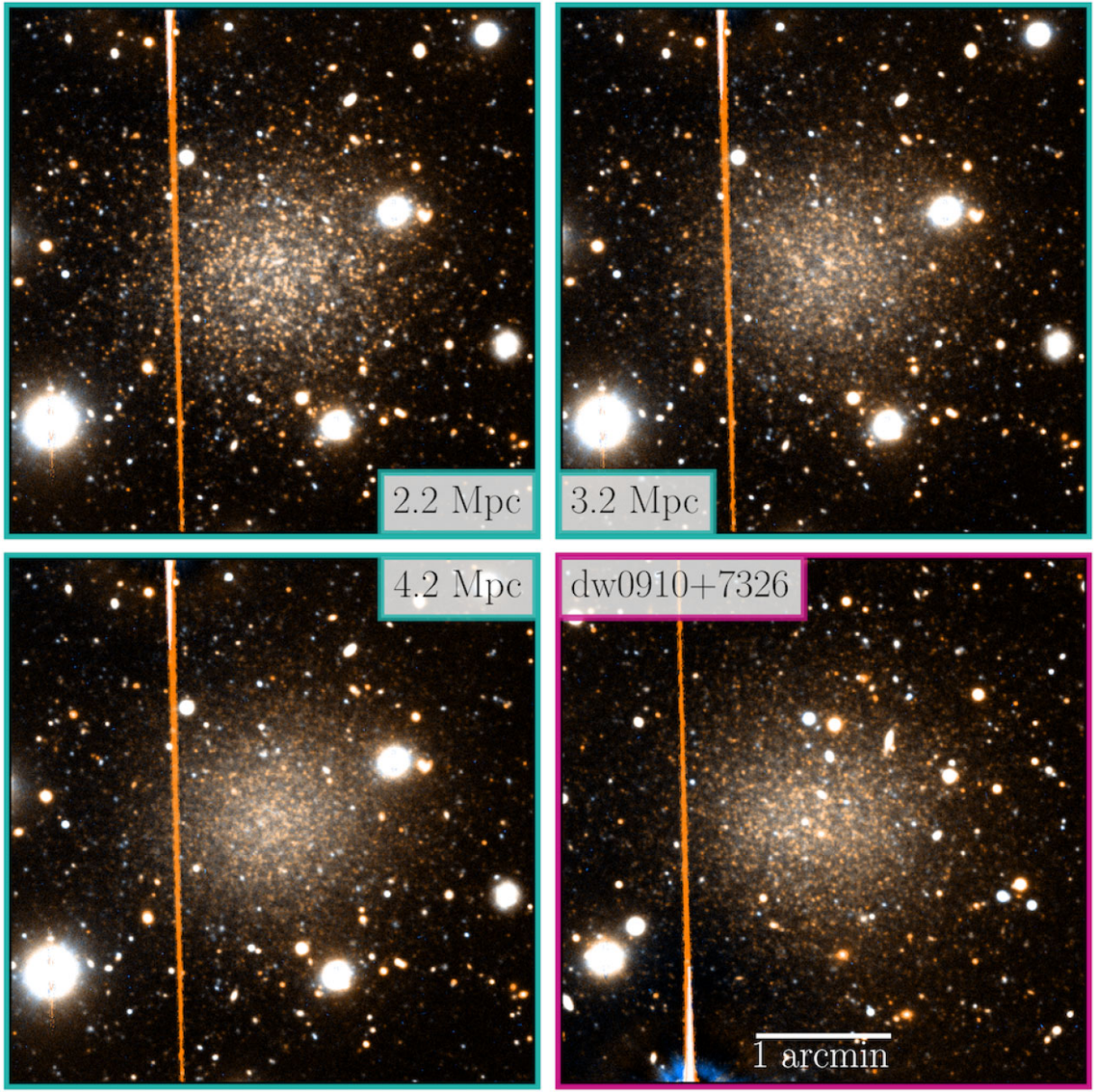

**Figure 2.** RGB images of dw0910+7326 (Blobby) and `ArtPop` (Greco & Danieli 2021) models at various distances generated using *R*- and *B*-band images, with the average of the two bands used for the green channel. The `ArtPop` models are fixed with the same structural and stellar population parameters, but the physical radii and masses are adjusted to ensure that the final image has the same angular size and apparent *R*- and *B*-band magnitudes as Blobby. Note that each mock galaxy and Blobby itself are semiresolved, albeit to different degrees; diffuse light modelling measures the integrated colour, but there is also a dominant (in terms of luminosity) semiresolved red giant population. Comparing the SBF-by-eye of the `ArtPop` models and Blobby further reinforces the final distance measured using the SBF method. The model at 2.2 Mpc is 'lumpier' than Blobby, while the model at 4.2 Mpc is smoother. The bright star shown in Fig. A1 has been subtracted in these images. Upper left: `ArtPop` model at 2.2 Mpc. Upper right: `ArtPop` model at 3.2 Mpc, which is the approximate measured distance of Blobby. Lower left: `ArtPop` model at 4.2 Mpc. Lower right: Blobby.

Observations were conducted on 2019 December 20. We took 10 × 300 s exposures in each filter. The seeing was ∼1.4–1.5 arcsec, which demonstrates again one of the strengths of the SBF method. Other distance measurements, such as TRGB analyses, require much better seeing, typically below ∼1.0 arcsec for objects at similar distances. A semirandom dither was used between each exposure on the order of the approximate size of the dwarf to ensure that the galaxy occupied a different location on the CCD in each exposure. This dithering pattern minimized the impact of bad pixels in the CCD and improved our ability to accurately measure the sky background. Each exposure contains 4608 × 2048 pixels with a pixel scale of 0.224 arcsec pixel$^{-1}$, for a total approximate area of 17.2 × 7.6 arcmin. Cut-outs of about 4 × 4 arcmin were created for the measurements and tests described in Section 3. For reference, Blobby has an approximate effective radius of 0.85 arcmin. *B*- and *R*-band magnitudes referenced in this work are in the Vega system; CFHT *g*- and *i*-band magnitudes use the AB system (Oke & Gunn 1983).

We developed an image processing pipeline for our work with LBT in order to improve processing of low surface brightness features in the final coadds. The individual exposures were overscan and bias subtracted. Master flats were constructed using ccdproc's COMBINER (Craig et al. 2017) with a median combine on twilight flats and stars masked. Each image was then registered to a common frame using the python toolkit included with *Astrometry.net*[3] (Lang et al. 2010). The zero-points and colour term corrections were determined by comparing stars in the image with the same stars in the Pan-STARRS catalogue (Chambers et al. 2016; Flewelling et al. 2020; Magnier et al. 2020; Waters et al. 2020). Galactic extinction corrections were determined from Schlafly & Finkbeiner (2011) using the IRSA GALACTIC DUST REDDENING AND EXTINCTION[4] (IRSA 2022) search tool.

There is a $m_R \sim 9$ star approximately 2 arcmin from Blobby's centre that heavily contaminated the background modelling, even when generously masked (see Fig. A1). For that reason, extensive tests were completed to determine how best to model and subtract the star and its diffuse halo from the image before the background was determined. Ultimately, the star was modelled with six stacked Gaussians in each band using pymfit, a python wrapper for IMFIT (Erwin 2015), and then subtracted from each exposure. Afterward, each exposure was background subtracted using a tilted plane fit of a *Source Extractor*-generated background (Bertin & Arnouts 1996). *Source Extractor* used a box size of 800 pixels and a custom mask, covering Blobby and the brightest foreground and background objects.

The final coadds in each filter were created from clipped means of the individual exposures, which were weighted according to the measured background levels. The PSF in each band was computed from a median stack of an unsaturated star in each exposure. SBF modelling and IMFIT results were not affected significantly by the choice of star for the PSF. The final star- and background-subtracted coadd cut-out of Blobby is shown in the bottom-right panel of Fig. 2. The RGB cut-out is made using a combination of the *R*- and *B*-band images. The green channel of the image is generated from an average of the two bands. The remaining panels of Fig. 2 show artificial galaxies generated using ArtPop (Greco & Danieli 2021) injected at different distances for comparison with Blobby's semiresolved stellar population. The ArtPop models and tests are described further in Section 3.

[3] https://github.com/dstndstn/astrometry.net
[4] https://irsa.ipac.caltech.edu/applications/DUST/

## 3 MEASUREMENTS AND VALIDATION

It is crucial to accurately measure Blobby's structural properties and SBF to put its local environment in context. Throughout this work we used mock galaxies created with ArtPop injected into the image to develop and validate the procedures ultimately used on the real dwarf. The properties of the mock galaxies are described in Section 3.1.

The SBF measurement procedure we used followed the power spectrum method described in Greco et al. (2021). First, the diffuse light from the galaxy was subtracted from the image. In this work, IMFIT was used to model the structure and diffuse light distribution of the galaxy (Section 3.2). We assumed that Blobby's structure could be accurately modelled by a single Sersic distribution. After the structure was measured, the next step was to normalize the scale of the SBF across the image by dividing the residual by the square root of the diffuse light model. Any sources of SBF not due to variations in positions of stars from the galaxy were masked. We used an absolute magnitude cut-off for objects detected in the residual to estimate which of the objects were likely associated with the galaxy (see Section 3.3). Finally, the masked and normalized residual was Fourier transformed and the azimuthally averaged power spectrum, $P(k)$, was computed:

$$P(k) = \sigma^2_{\text{SBF}} \times E(k) + \sigma^2_{\text{WN}}. \qquad (2)$$

The power spectrum is a function of the spatial frequency, $k$, in units of inverse pixels and can be modelled as a combination of the SBF signal, $\sigma^2_{\text{SBF}}$, a constant white noise signal, $\sigma^2_{\text{WN}}$, and the expectation power spectrum, $E(k)$ (Tonry et al. 1990), which is a convolution of the PSF power spectrum and the mask. For our purposes, we can measure $P(k)$, $E(k)$, and $\sigma^2_{\text{WN}}$ directly to extract $\sigma^2_{\text{SBF}}$. The SBF calculation is described in Section 3.3.

We translate from the measured $\sigma^2_{\text{SBF}}$ to a distance using an intermediate variable, the SBF magnitude. The apparent SBF magnitude was computed using

$$m_{\text{SBF}} = m_{\text{zpt}} - 2.5 \log \left( \frac{\sigma^2_{\text{SFB}} - \sigma^2_r}{N_{\text{pix}}} \right), \qquad (3)$$

following methods described in Greco et al. (2021) and Carlsten et al. (2019). Recall that $\sigma^2_{\text{SBF}} \equiv \bar{L}$ has units of luminosity, which is why it can be used to calculate a magnitude (and ultimately a distance) in this way. Here, $m_{\text{zpt}}$ is the photometric zero-point and $N_{\text{pix}}$ is the number of unmasked pixels in the residual image. The SBF signal from the galaxy, $\sigma^2_{\text{SBF}}$, was computed by fitting equation (2), and is dependent on the luminosity function of the stellar population as described in equation (1). The residual variance, $\sigma^2_r$, is the SBF signal from contaminating sources. We estimated the residual variance by measuring the SBF signal in background regions of the image (Section 3.3).

The final step necessary to measure a distance was to use an SBF magnitude–colour relation to calculate an *absolute* SBF magnitude, $M_{\text{SBF}}$, from Blobby's measured colour. The SBF magnitude–colour relation holds because the SBF signal is dependent on the stellar population, which affects the integrated colour of the galaxy. There are several available calibrations, some using empirical relations and others using theoretical isochrones. In this work, we use the empirical *g–i* relation from Carlsten et al. (2019) along with $B - R$ colour transformations from MIST (Choi et al. 2016; Dotter 2016). The distance modulus, $\mu = m_{SBF} - M_{SBF}$, follows directly from the measured absolute and apparent SBF magnitudes and can be used to calculate a distance, $D$, using the usual method,

$$\begin{aligned}D &= 10^{\frac{1}{5}(\mu+5)} \\ &= 10^{\frac{1}{5}(m_{SBF}-M_{SBF}+5)}.\end{aligned} \quad (4)$$

Repeated injection and recovery of the structure, SBF, and distances of mock ArtPop galaxies demonstrated the biases and scatter to be expected for the real dwarf measurement (Section 3.4).

### 3.1 Artificial galaxy comparisons

To develop and validate the procedures described in this section, mock galaxies made using ArtPop were injected into the image. ArtPop builds galaxies star by star, based on given stellar distribution and stellar population parameters (Greco & Danieli 2021). Since the mock galaxies' structural parameters and SBF were known exactly, they could be used to test the robustness of the modelling.

For this work we used the *UBVRIplus* photometric system with a Kroupa initial mass function (Kroupa 2001) and a simple stellar population (SSP) to generate ArtPop models. The positions of the stars in the models were based on a Sersic distribution which matched measurements of Blobby's structure. Fig. 2 demonstrates the appearance of ArtPop models placed at different distances in the image and their similarity to Blobby. The SBF-by-eye of the mock galaxies is readily apparent with the image variance smoothing out in more distant models. The mock galaxy with SBF-by-eye most similar to Blobby (upper right) is also the one injected closest to Blobby's measured distance.

The mock ArtPop galaxies used to test the structure and SBF measurement procedures were built using MIST isochrones (Choi et al. 2016; Dotter 2016) and were used to constrain properties of the stellar population for the real galaxy. Different choices for the metallicity, age, and stellar mass of each ArtPop model injected at a given distance resulted in mock galaxies with different colours and apparent magnitudes. By matching the properties of the mock galaxies to the real dwarf we could both validate our measurement procedures for Blobby and make predictions about its likely stellar population.

There are degeneracies between different SSP parameters that each reproduce mock galaxies resembling Blobby. For instance, different combinations of age and metallicity result in the same colour for a given stellar population. Without more detailed metallicity information, it is impossible to say for certain what the age of the real stellar population must be, and vice versa. Fig. 3 shows the range of resulting mock galaxy colours given different ages and metallicities for the underlying SSP. The red star marks the metallicity and age of the ArtPop models used to test and validate the procedures in this work. The measured colour of Blobby along with the 1$\sigma$ uncertainty in the colour are marked by the white solid and dashed lines, respectively. The vertical dashed black lines mark the upper and lower bound in metallicity predicted from the mass–metallicity relation (MZR) for dwarf galaxies in Kirby et al. (2013) for a galaxy with the upper and lower mass limits of Blobby. The vertical black dashed lines also include the effect of the RMS about the best-fitting MZR. The upper and lower mass limits (shown in Table 1) are determined by taking the statistical uncertainty in the measured SBF distance (see Section 4), and finding the required mock galaxy mass for those limits to match the apparent magnitudes of the real galaxy. Assuming the SSP reasonably represents the stellar population of Blobby, Fig. 3 indicates that Blobby likely has a relatively metal-rich stellar population compared to the typical dwarf galaxy of its mass. If we assume that Blobby's metallicity does not deviate dramatically from that expected from this type of dwarf galaxy, then it is also clear that the age of the stellar population is old ($\gtrsim$10 Gyr). Interestingly, the age we estimate for Blobby's stellar population is similar to the 90 per cent star formation time-scales for two low-mass M81 satellites in Smercina et al. (2022), indicating a potentially exciting avenue for future study. The particular stellar population parameters used for the testing described in this section were chosen to reasonably match the expected metallicity from Kirby et al. (2013) while still matching the apparent magnitudes from preliminary measurements of the real galaxy. We argue that the assumption of an SSP is reasonable because SBF magnitude–colour relations for SSPs are in excellent agreement with observations for galaxies with $0.5 \lesssim g - i \lesssim 0.9$ (Cantiello et al. 2018; Greco et al. 2021). Using colour conversions from MIST, we find that Blobby has $g - i \sim 0.78$.

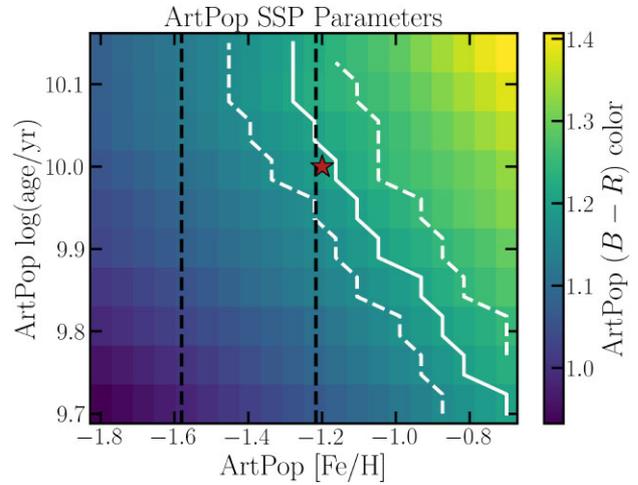

**Figure 3.** Demonstration of the $B - R$ colours of a simple stellar population with given age and metallicity parameters in ArtPop. The white solid line marks the bins with colours closest to Blobby, and the white dashed lines show the 1$\sigma$ spread in colour expected based on measured colour error tests with ArtPop models. The black dashed lines show the expected range in metallicities from the mass–metallicity relation (MZR) of Kirby et al. (2013). The metallicity limits were determined using the RMS of the MZR and the limits on Blobby's mass. Mass limits are those required to keep the apparent magnitudes of the ArtPop models consistent with Blobby's given the statistical uncertainty in Blobby's distance (see Section 4). The star shows the age and metallicity used for the ArtPop models in the tests described in Section 3.

### 3.2 Structural parameters

We used a python wrapper for IMFIT called pymfit[5] to measure the structural properties of the diffuse light of the real and mock galaxies in both *B*- and *R*-band images. IMFIT requires several inputs, including the background-subtracted image of the galaxy, the image mask, the noise level in the image, the PSF, and the details of the model to be fit. We created the mask in each band by hand in combination with pymfit's masking procedures, which depends on SEP (a python wrapper for *Source Extractor*; Bertin & Arnouts 1996; Barbary 2016) to detect bright background and foreground objects. After object detection, pymfit convolves the mask with a Gaussian kernel to increase the area around each detection in order to minimize the impact of diffuse light around background and foreground objects. The masks in both bands were then inspected by eye and additional masks were added to bright foreground stars where necessary, again to limit the impact of diffuse light on the IMFIT measurement of the galaxy. Finally, the *B*- and *R*-band masks were combined into the final mask shown in purple in Fig. 4. Special care was taken with the $m_R \sim 9$ star removed in Fig. 2 but shown in Fig. A1. All IMFIT modelling was conducted on images in which the star had been modelled and removed. A flat variance was provided along with the PSF described in Section 2. We conducted tests providing IMFIT with the sky level from the background subtraction modelling instead of the flat variance image, but the results were not changed significantly.

---

[5] https://github.com/johnnygreco/pymfit

**Table 1.** Properties of dw0910+7326 (Blobby). The first section of the table indicates bias-corrected measured quantities, 1σ statistical uncertainties, and biases (measured from ArtPop tests; Section 3.4) used to correct to the measured quantities. The surface brightnesses were calculated using relations from Graham & Driver (2005). The position angle (PA) is measured counterclockwise from the positive y-axis. Since the ArtPop models are not located in the same position as Blobby relative to the star, which likely has an effect on the error in PA and since Blobby's ellipticity is very low, the bias and statistical error on this measurement were simply combined. The second section of the table indicates measured quantities that rely on the SBF magnitude–colour relation and includes both statistical uncertainties (similar to the first section – listed first) and systematic uncertainties (listed second) resulting from the chosen SBF magnitude–colour relation (Carlsten et al. 2019). The uncertainties on $D_{M81}$ were computed from the line-of-sight 1σ statistical and systematic uncertainties listed for $D$ and the 1σ spread in TRGB distances to M81 added in quadrature. The third section of the table includes an estimated stellar mass based on magnitude comparisons between Blobby and ArtPop models at a range of distances chosen given the statistical and systematic distance errors, respectively.

| Parameter | Bias-corrected value | Bias |
|---|---|---|
| RA (hh:mm:ss) | 09:10:13.45 | – |
| Dec. (dd:mm:ss) | +73:26:19.15 | – |
| $D_{M81,\,proj}$ | 5.67 deg | – |
| $\mu_{0, R}$ | 24.71 ± 0.60 mag arcsec$^{-2}$ | – |
| $\mu_{e, R}$ | 25.79 ± 0.60 mag arcsec$^{-2}$ | – |
| $\mu_{0, B}$ | 25.92 ± 0.60 mag arcsec$^{-2}$ | – |
| $\mu_{e, B}$ | 26.99 ± 0.60 mag arcsec$^{-2}$ | – |
| $r_e$ | 50.8 ± 1.4 arcsec | −0.51 |
| $m_R$ | 14.76 ± 0.05 mag | −0.125 |
| $m_B$ | 15.96 ± 0.06 mag | −0.038 |
| $B - R$ | 1.20 ± 0.04 mag | 0.088 |
| $m_{SBF, R}$ | 26.34 ± 0.07 mag | 0.052 |
| n | 0.65 ± 0.02 | 0.011 |
| Ellipticity | 0.10 ± 0.01 | −0.013 |
| Position angle | 85.1 ± 15 deg | – |
| GALEX $m_{NUV}$ | >18.9 mag | – |
| GALEX $m_{FUV}$ | >17.7 mag | – |
| $r_e$ | $790^{+28+101}_{-28-89}$ pc | – |
| $M_R$ | −12.78 ± 0.07 ± 0.25 mag | – |
| $M_B$ | −11.57 ± 0.08 ± 0.25 mag | – |
| $M_{SBF, R}$ | −1.19 ± 0.04 ± 0.26 mag | – |
| $D$ | $3.21^{+0.15+0.41}_{-0.15-0.36}$ Mpc | – |
| $D_{M81}$ | $587^{+379}_{-222}$ kpc | – |
| SFR | <2.7 × 10$^{-4}$ M$_\odot$ yr$^{-1}$ | – |
| $M_\star$ | (9.5 ± 1.0 ± 2.2) × 10$^6$ M$_\odot$ | – |

Several options were tested to determine the best combination of input model functions and parameters to measure the galaxy's structure. We tried a simple Sersic alone and in combination with both a flat plane and a tilted plane. The rationale for including a plane in the fit was to help mitigate any oversubtraction or undersubtraction resulting from the background subtraction step. Based on repeated tests with injecting and then recovering the structure of ArtPop-generated artificial galaxies in the image (see Section 3.4), the final set of model functions used for measuring the structure of Blobby were a tilted plane and a Sersic profile. The same tests also determined that the final magnitudes of the mock galaxies were recovered most accurately when the structure in both *B*- and *R*-band images was determined separately rather than having the structure of one band dependent on measurements in the other. In Table 1, the structural parameters are taken from the *B*-band results, while the colour and apparent magnitudes are determined from each band's respective measurement, corrected using the biases listed in Table 1.

Fig. 4 demonstrates the steps of the IMFIT modelling on *R*-band images of both Blobby (top row) and a mock galaxy (bottom row) injected near the same bright star. The left panels show the original images along with the masked regions in purple. Notice the large masked region where the bright star was located before the star subtraction was implemented. The modelled diffuse light from each galaxy is shown in the middle panels and includes both the Sersic and tilted plane profiles. The difference between the original images and models results in the residual images on the right. The ArtPop model in the bottom row has the same measured structure as the final Blobby measurement and was injected at the measured distance. There is no significant large-scale residual structure in the right panels, so we argue that the assumption of a Sersic distribution is fair.

### 3.3 Surface brightness fluctuations

The measured SBF is influenced heavily by the quality of the IMFIT modelling of the galaxy. SBF are sensitive to large-scale deviations in the light distribution of the dwarf. If the diffuse light in the outskirts of the dwarf is underestimated, for instance, it will lead to large-scale fluctuations in the light profile of the residual, which will be imprinted on the measured power spectrum. The same argument makes accurate background subtraction particularly crucial and that is why including the tilted plane in the IMFIT modelling also helps with the SBF measurement. Since the SBF signal in *R*- band is more pronounced than in *B*-band, we used the *R*-band IMFIT results for the SBF measurement.

The final SBF measurements of Blobby and the mock galaxies used the power spectrum method introduced in the preamble of Section 3. In each of these measurements, the residual image was created by subtracting both the tilted plane and the Sersic distribution computed using IMFIT from the original image (Fig. 4) as described in Section 3.2. Afterward, the residual was normalized using only the Sersic component of the model, so that the scale of SBF is constant across the image. In order to remove any contamination of the SBF signal from background sources, we mask all sources brighter than $M_R = -5$ at the distance of 3.0 Mpc. The $M_R = -5$ absolute magnitude threshold was chosen to reduce contamination from globular clusters while leaving RGB stars in the galaxy unmasked, as described in Carlsten et al. (2019). Because the absolute magnitude cut assumes a distance, we began our tests with a benchmark distance of 3.0 Mpc based on preliminary Blobby measurements and because similarly distant ArtPop models looked qualitatively similar to the real galaxy (Fig. 2). As in the IMFIT masking, the SBF mask was convolved with a Gaussian kernel to increase the size of the mask and ensure that any diffuse light from contaminating sources was covered. The masked residual (shown for Blobby in the inset image of Fig. 5) was then Fourier transformed and the azimuthally averaged power spectrum was computed.

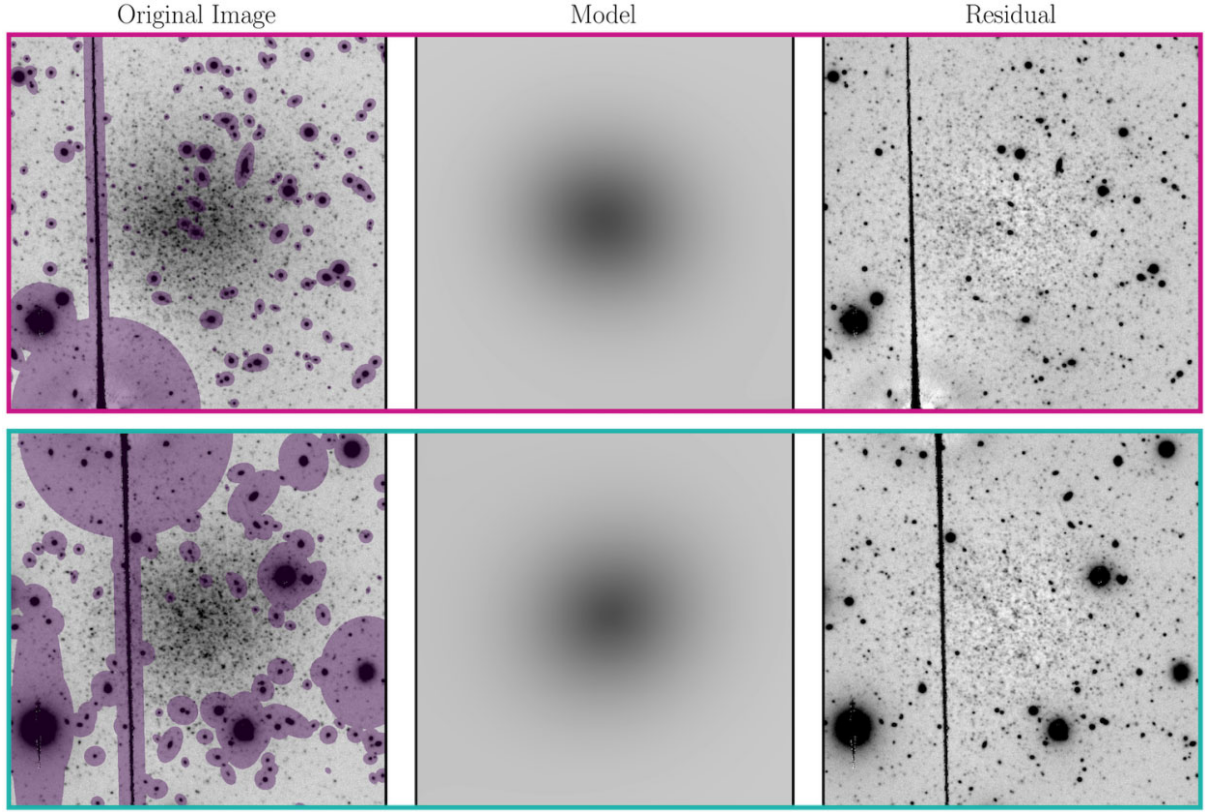

**Figure 4.** Top: *R*-band IMFIT modelling of Blobby. Bottom: *R*-band IMFIT modelling of an `ArtPop` model with properties meant to mimic Blobby. Left: Original *R*-band image of (faux) Blobby where the purple shaded regions show the areas masked for IMFIT modelling. Centre: IMFIT model (tilted plane + Sersic). Right: Residual image where the model has been subtracted from the original.

The black line in Fig. 5 shows the observed power spectrum, $P(k)$ in equation (2). The purple line is the fit to the observed power spectrum, and the diagonal dashed grey line is $E(k)$, the expectation power spectrum. $E(k)$ is measured first by zero-padding the PSF as in Greco et al. (2021) and using NUMPY's `fft` tools to compute the power spectrum for both the PSF and the mask. The PSF and mask power spectra are convolved using ASTROPY's `convolve_fft` function and then azimuthally averaged. The horizontal dashed grey line shows the level of the white noise, $\sigma_{WN}^2$, which has been normalized to 1. For Blobby and for each mock galaxy, we repeated the SBF measurement 100 times with random variations in the upper and lower $k$ limits. We computed the median value of these 100 iterations to get the final SBF signal. As mentioned previously, the resulting $m_{SBF}$ is sensitive to changes in the minimum $k$ limit (small $k$ in Fig. 5), corresponding to large scales in the image. Imperfect background subtraction or Sersic model fitting can easily result in structures on the scale of the image. This was especially concerning because of the bright star so close to Blobby. To mitigate the effect of these structures we increased the typical range of lower $k$ limits allowed in the iterations to 0.04–0.1. We confirmed using the `Artpop` model tests that despite the increased lower $k$ limit, we could still reliably recover the correct apparent SBF magnitude of the mock galaxies. The upper $k$ limit we used was 0.3–0.4.

As in Carlsten et al. (2019), we empirically measure the contamination from sources unrelated to the dwarf by repeating the SBF procedure in blank fields selected throughout the image. These contaminating sources include contributions to the SBF signal from background galaxies and correlated pixels from the image registration step. In total, we hand-selected eight blank fields, avoiding foreground MW stars and bright background galaxies. In each iteration of the SBF measurement, a random blank field was selected, and the SBF of the blank field ($\sigma_r^2$ in equation 3) was subtracted from the galaxy signal. The blank fields were normalized and masked in the same way as the galaxy. The tan lines in Fig. 5 show the measured power spectra from each of the blank fields. The blank field corrections do not account for any contributions from globular clusters, but as with many similar galaxies, the expected number of globular clusters is small, likely zero (Forbes et al. 2018). If there are globular clusters then the $M_R = -5$ limit used in the masking procedure should remove them given the peak of the globular cluster luminosity function ($M_I \sim -8$; Miller & Lotz 2007). It is not impossible that globular clusters exist beyond our masking threshold but it is unlikely, so we neglect the effect of globular clusters for the purposes of this paper.

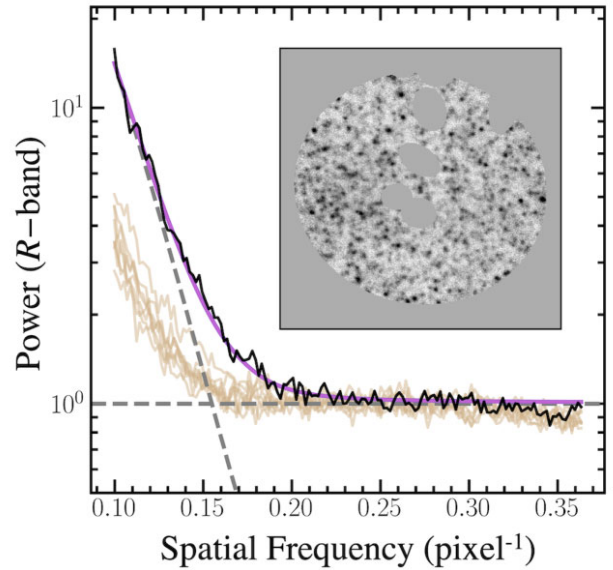

**Figure 5.** SBF measurement for Blobby. The black line shows the measured power spectrum, the purple line shows the fit to the power spectrum, the tan lines show the signal from the eight blank fields, and the grey dashed lines show the PSF power spectrum and the white noise; all have been normalized by the same amount such that the white noise level is at 1.0. We measure an S/N ∼13 for Blobby. Inset image: Masked and normalized residual image used for SBF measurement.

In order to recover the distance, $D$ in equation (4), from the apparent SBF magnitude, $m_{SBF}$, we used a calibration that relates the galaxy's integrated colour to an absolute SBF magnitude, $M_{SBF}$. Empirical calibrations of this kind generally measure the absolute SBF magnitude of galaxies with well-understood distances from reliable methods like TRGB analyses. Purely theoretical calibrations derive the absolute SBF magnitudes and colours using stellar population synthesis models, and semi-empirical calibrations use some combination of both theoretical and empirical methods. There are several available calibrations for galaxies similar to Blobby, including semi-empirical $M_{SBF,R}$–$(B - R)$ relations for dwarf elliptical galaxies calculated in Jerjen et al. (2000) and then revised in Jerjen et al. (2004), the empirical Carlsten et al. (2019) $M_{SBF,i}$–$(g - i)$ relation, the related Carlsten et al. (2021b) $M_{SBF,r}$–$(g - r)$ relation, the empirical Cantiello et al. (2018) $M_{SBF,i}$–$(g - i)$ relation, and the theoretical calibrations from MIST (Choi et al. 2016; Dotter 2016), among others.

For empirical and semi-empirical relations, the colour range of galaxies used for the calibration differs, and extrapolations to bluer or redder colours can lead to increased uncertainties. The Cantiello et al. (2018) $(g - i)$ relation, for example, is valid for galaxies in the approximate range $0.82 \lesssim g - i \lesssim 1.05$, whereas the Carlsten et al. (2019) relation covers bluer galaxies in the range $0.3 \lesssim g - i \lesssim 0.8$. In order to compare the results of these relations for Blobby, we used MIST to generate conversions between the $B - R$ colours and the relevant colour of each calibration, as well as conversions between absolute SBF magnitudes in different bands. We did this by generating MIST isochrones of SSPs across a wide range of age and metallicity space using the relevant photometric systems (e.g. *UBVRIplus* or *CFHTugriz*). The colours and absolute SBF magnitudes of these otherwise identical isochrones were 'measured' and compared such that for any $B - R$ colour, we could use any calibration to determine an $R$-band SBF magnitude. In this way, we also estimated the $g - i \sim 0.78$ colour of Blobby, for example, which is within the range of the Carlsten et al. (2019) calibration and a slight extrapolation from the Cantiello et al. (2018) calibration.

Each calibration mentioned above results in slightly different absolute SBF magnitudes for a given $B - R$ colour, and the difference between calibrations changes significantly depending on the colour in question. The calibrations roughly converge in the colour range Blobby inhabits. We determined that the Jerjen et al. (2000) and Jerjen et al. (2004) relations resulted in larger distance errors for the ArtPop models we considered. The discrepancy is likely due to the use of older theoretical models from Worthey (1994) and Bertelli et al. (1994). Thus, we focused on the newer empirical calibrations in Fig. 6, which shows the relative differences between distance results using each calibration in the left panel. The Carlsten et al. (2019) and Cantiello et al. (2018) relations performed best in mock galaxy testing (Section 3.4), and since Blobby's measured $g - i$ colour corresponds best with the Carlsten et al. (2019) calibration, that is the calibration we use for our final result (Section 4).

### 3.4 Validation

To improve our understanding of measurement biases and uncertainties, we injected ArtPop models in the image and recovered them in the same way as we would for the real galaxy. There were two regions in the image selected for the ArtPop model testing, one as far from the bright star and other contamination as possible (the 'ideal' location) and one very close to the star in a similar position to Blobby (the 'realistic' location; Fig. 2). The ideal and realistic locations were especially important for understanding the extent to which the bright star (Fig. A1) would skew the results of the IMFIT and SBF modelling. Initial testing, for instance, did not include the star subtraction step described in Section 2, which revealed clear issues with background oversubtraction in the ideal location and contamination from the star in $R$- band in the realistic location. As a result, we implemented the star subtraction step of the pipeline before the background was estimated. This improved the background subtraction and magnitude measurements in the ideal location significantly. In the realistic location, $B$-band magnitude recovery also improved, although a bias in the $R$-band magnitude persisted. Understanding the severity and direction of the errors imposed by the star allowed us to correct for the star's effect on our final distance measurement of Blobby. Since Blobby's position is more similar to the realistic ArtPop location, the scatter, and biases shown in Table 1, those used to correct Blobby's final measurements, and those quoted throughout this work are those derived from the realistic ArtPop tests only.

In order to mimic the IMFIT and SBF modelling of the real galaxy as closely as possible, efforts were made to create ArtPop models that closely resembled the real galaxy. For instance, a preliminary measurement was made on the real galaxy so that the resulting structural parameters could be used in the ArtPop models. Once the IMFIT modelling of the ArtPop models was improved, the structure of Blobby was measured again and applied to the ArtPop models to further improve the match. In order to test the IMFIT and SBF modelling procedures on galaxies at different distances, tests included ArtPop models at every 500 kpc increment from 2.0 to 4.5 Mpc. The masses of these mock galaxies were adjusted for each distance in order to closely match the $R$- and $B$-band apparent magnitudes of Blobby. In total we used 2400 uniquely generated models to test the effects of different positions in the image, distances to the galaxies, and background subtraction procedures. A final sample of 600 ArtPop models was used to obtain the biases and scatter cited in Table 1 (100 at each distance increment) with the same background subtraction and realistic location in the image as Blobby.

The SBF and IMFIT results on the mock galaxies were examined using the measured distribution of errors for each parameter, $X$, where $X_{error} = X_{true} - X_{measured}$, except in the case of the radius, where the fractional error was used instead. All distributions appeared roughly Gaussian. The biases quoted in Table 1 are calculated simply from the median of each error distribution, and the statistical uncertainties for those parameters were determined from the $1\sigma$ spread in the distributions. For the radius, the fractional error was used to calculate the corresponding bias and uncertainty in arcseconds cited in Table 1. The systematic uncertainties listed in Table 1 result from the Carlsten et al. (2019) SBF magnitude–colour calibration. We argue that this simple approach to calculating parameter statistical uncertainties is justified because systematic uncertainties from the SBF magnitude–colour calibration dominate the distance uncertainty budget. There were no significant differences in any of the parameter biases or spreads between models of different distances, except in the case of $m_{\rm SBF, R}$ which is discussed below. Similarly, the background subtraction method did not significantly affect the error distributions, although the small difference in recovered distances is included as a statistical uncertainty in the final distance, $D$, listed in Table 1. The final results for the parameter values in Table 1 have been corrected by the measured biases so that the error distributions are centred at zero. All other derived parameters (those without biases listed) are calculated using the bias-corrected values.

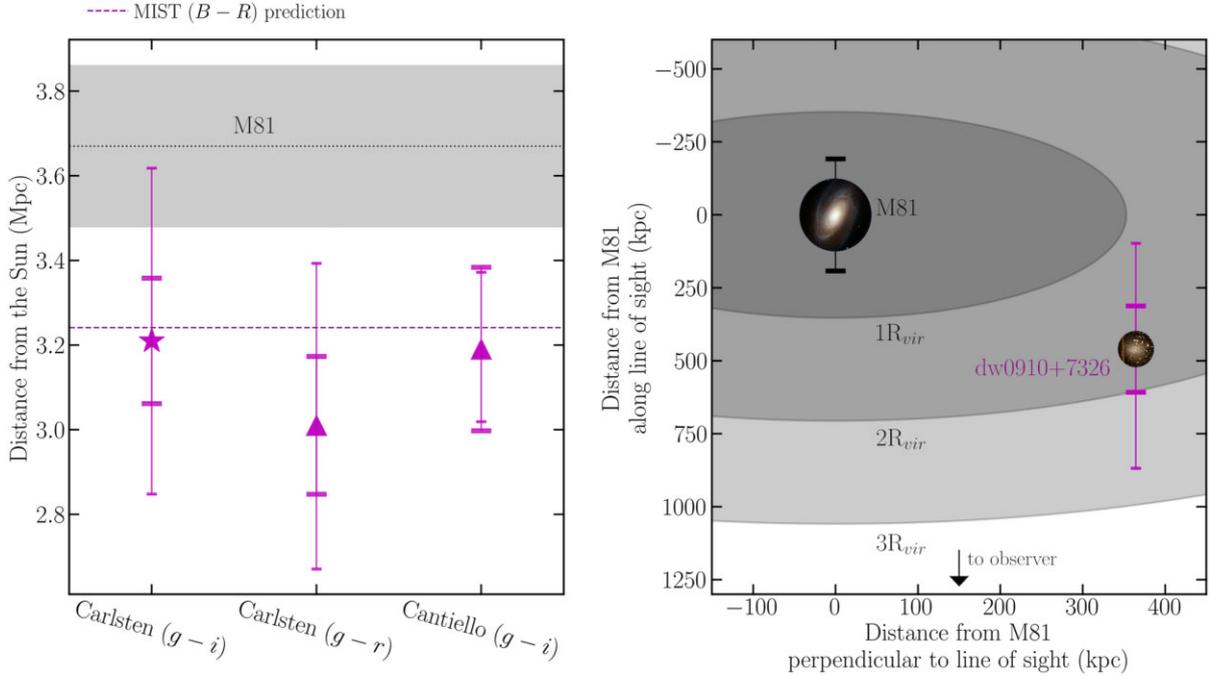

**Figure 6.** Left: The distances measured for Blobby using different SBF magnitude–colour calibrations. The star indicates the calibration that was used for the final result, shown in the plot on the right and included in Table 1. The error bars show the effect of the statistical uncertainty (thick caps) and the systematic uncertainty (thin caps). The magenta dashed line shows the distance calculated using the absolute SBF magnitude–$(B - R)$ colour relation from SSPs generated with MIST (Choi et al. 2016; Dotter 2016). The black dotted line shows the mean of the TRGB distance measurements for M81 listed in NED, and the shaded region indicates the $1\sigma$ spread in the M81 TRGB distance measurements. References for the distance calibrations in the order listed on the x-axis: Carlsten et al. (2019), Carlsten et al. (2021b), and Cantiello et al. (2018). Blobby's calculated $g - i \sim 0.78$ colour is an extrapolation blueward of the galaxies considered in Cantiello et al. (2018), which is why we do not use this calibration in the right panel or for the final distance measurement in Table 1. Right: Top-down view of the plane containing M81, Blobby, and the Sun (images of M81 and Blobby are not to scale), where M81 is set at the origin. The error bars for M81 show the $1\sigma$ spread in TRGB distances to M81 in NED and the error bars for Blobby are the same as those on the star point in the left plot. The shaded regions show the location of $1R_{\rm vir}$, $2R_{\rm vir}$, and $3R_{\rm vir}$ for the combined M81–M82 system assuming a distance of 3.67 Mpc. M81 image credit: NASA, ESA, and the Hubble Heritage Team (STScI/AURA). Acknowledgment: A. Zezas and J. Huchra (Harvard-Smithsonian Center for Astrophysics)

In the course of testing, we discovered a small but ultimately negligible bias in the apparent SBF magnitude measurement as a function of distance. The trend in $m_{\rm SBF, error}$ was roughly linear from about 0.08 mag for the closest models and 0.01 mag for the farthest models. To investigate the effect of this trend on the recovered distance, we took the median bias of *all* 600 ArtPop models in the final sample regardless of distance and used it to correct the measured SBF magnitude of ArtPop galaxies at specific distances. We compared the resulting distances to those calculated assuming that the SBF magnitudes and colours had been measured perfectly (using the known SBF magnitudes and colours of the models). Using the 3.0 Mpc models, for example, the 'perfect' and 'measured' distances differed by less than 1 per cent for all calibrations shown in Fig. 6. We repeated this measurement for the models at the extremes of the distances we explored and we got similarly negligible distance errors. Therefore, we conclude that the effect of the bias is negligible compared to the uncertainties.

## 4 RESULTS

Following confirmation of our modelling procedure with ArtPop, we analysed Blobby itself using the same methods. Table 1 summarizes the final results of the IMFIT and SBF modelling, all of which have been corrected by the measured biases using the respective parameters of the 600 ArtPop models described in Section 3.4. As mentioned in Section 3, Blobby's structure was measured using a Sersic model. We argue that because there is no significant structure in the residual image shown in the top right panel of Fig. 4, a single Sersic function is sufficient for this galaxy. The amount of residual structure is similar to the remaining structure in the residual of the ArtPop model, which used a Sersic function to determine the positions of the stars in the mock galaxy.

The final power spectrum fit for Blobby is shown as the purple line in Fig. 5, and the SBF masked residual is shown in the inset image. We evaluated the S/N of the measurement using the method described in Carlsten et al. (2019). The signal was measured from the median of the 100 iterations of the power spectrum fit corrected by the relevant blank field fit as described in Section 3.3. The noise was estimated from the standard deviation of the signal measurements. We measured an S/N ∼13, a factor of 6.5 higher than the S/N > 2 cut-off Carlsten et al. (2019) employed to discard unreliable measurements.

The final distance result for Blobby in the context of M81 is summarized in Fig. 6. The left panel includes each SBF magnitude–colour calibration we considered, where the chosen calibration (the $g − i$ Carlsten et al. 2019 relation) is shown as a magenta star. This calibration was chosen because it consistently resulted in the least error in ArtPop testing and because Blobby's estimated $g − i$ colour is included in the range of galaxy colours used to derive the calibration. We recovered a final distance of $3.21^{+0.15+0.41}_{-0.15-0.36}$ Mpc. The statistical uncertainties (listed first) include the $1\sigma$ spread in the recovered SBF magnitude error from ArtPop tests and the relative difference in recovered SBF magnitudes from different background subtraction techniques. The systematic uncertainties (listed second) were calculated from the residual RMS of the Carlsten et al. (2019) empirical calibration. This distance is compared to the $3.67 \pm 0.19$ Mpc distance to M81 (the black dotted line and shaded region in the left panel of Fig. 6). We argue that the measured distance makes sense in the context of Fig. 2 and is not likely in the outer limits that the error bars suggest considering the qualitative similarities between the real galaxy and the ArtPop model at 3.2 Mpc.

The right panel of Fig. 6 shows the implications of the measured SBF distance for Blobby's local environment. The uncertainties on the distances to both Blobby and M81 suggest that Blobby is likely between 1 and $2.7 R_{vir}$ with a central measured value of approximately 590 kpc or $1.7 R_{vir}$ from M81 (using the M81–M82 system virial radius of $R_{vir} \sim 353$ kpc), meaning Blobby lies in a region where simulations predict that the abundance of backsplash galaxies around similar mass hosts is roughly 50 per cent (Gill et al. 2005; Garrison-Kimmel et al. 2014; Simpson et al. 2018; Buck et al. 2019; Haggar et al. 2020; Applebaum et al. 2021; Bakels, Ludlow & Power 2021; Santos-Santos et al. 2023).

Blobby's appearance and red colour are consistent with a quenched stellar population. In order to better quantify Blobby's quiescence, we searched for a UV signal – a telltale sign of recent (≲100 Myr) star formation – in archival GALEX data (Leroy et al. 2019). We searched the Barbara A. Mikulski Archive for Space Telescopes (MAST) portal[6] and found two ∼100 s exposures, one in FUV and the other in NUV, from the All-Sky Imaging Survey (AIS) and one ∼1700 s NUV exposure from the Medium Imaging Survey (MIS). Blobby was located at the extreme outskirts of each of these images. Neither the AIS nor the MIS images showed any indication of a UV signal at Blobby's location beyond what appear to be image artefacts near the edge of the MIS image. We therefore argue that this is a non-detection and sought to estimate an upper limit on Blobby's SFR using methods like those in Davis et al. (2021). Using the NASA/IPAC Infrared Science Archive (IRSA)[7] (Team COSMOS 2020), which contains background-subtracted GALEX images, we took the image RMS from the nearest NUV and FUV fields to calculate $3\sigma$ lower limits on Blobby's UV flux within $1 R_{eff}$ (see Table 1). We used a relation between flux and SFR from Leroy et al. (2019) to calculate a $3\sigma$ upper limit of ≲$2.7 \times 10^{-4}$ $M_\odot$ yr$^{-1}$ on Blobby's SFR assuming a distance of 3.21 Mpc. In combination with Blobby's appearance and red colour, the low SFR indicates that Blobby is very likely quenched. Taken together with the predicted 3D distance, Blobby's quenched status disfavours the possibility that it is an unprocessed field galaxy. However, more information, like line-of-sight velocity, H I mass, or halo mass of Blobby, is required to rule out the field scenario (Buck et al. 2019).

[6] http://galex.stsci.edu
[7] https://irsa.ipac.caltech.edu

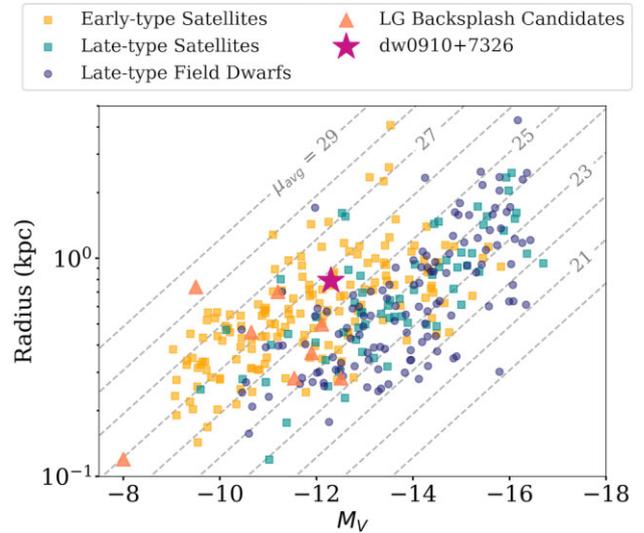

**Figure 7.** Size and absolute *V*-band magnitudes of Local Volume dwarf galaxies from Carlsten et al. (2021c) and Local Group backsplash candidates from Blaña et al. (2020) and Santos-Santos, Navarro & McConnachie (2023) with supplementary size and magnitude data from McConnachie (2012). A *V*-band magnitude was estimated for Blobby by generating an ArtPop model with Blobby's measured distance and *B*- and *R*-band magnitudes. Like other backsplash candidates, Blobby more closely resembles early-type satellite galaxies than late-type galaxies, even those that are in field environments.

We can compare Blobby to similar low-mass galaxies. We calculate a radius of about $790^{+28+101}_{-28-89}$ pc. When the mass of ArtPop models is tuned to match the apparent magnitudes of Blobby at 3.21 Mpc, we find an approximate stellar mass of $(9.5 \pm 1.0 \pm 2.2) \times 10^6$ $M_\odot$. This result puts Blobby above the (∼470 pc) best-fitting $R_{eff}$–log$M_\star$ relation for Local Group (LG) dwarfs from Danieli, van Dokkum & Conroy (2018), but within the $2\sigma$ scatter (∼920 pc). Blobby's size, although more extended than average, is consistent with other Local Group and Local Volume dwarf galaxies for its estimated mass. Fig. 7 shows the *V*-band absolute magnitude and effective radius of Local Volume dwarf galaxies from the sample in Carlsten et al. (2021c). The *V*-band magnitude for Blobby has been estimated from an ArtPop model with the same *B*- and *R*-band magnitudes as the real galaxy. Blobby appears to align more closely with the early-type satellite galaxy population than any of the late-type populations, even those late-type galaxies in more isolated field environments. The same phenomenon is apparent for the other candidate Local Group backsplash galaxies shown as orange triangles. In simulations, backsplash galaxies are much more similar to satellite galaxies than to field galaxies because of their interaction history (Benavides et al. 2022). Similarly, the measured absolute *B*-band magnitude of Blobby is typical of transitional and spheroidal dwarfs in Weisz et al. (2011b) but dimmer than the vast majority of dwarf irregular galaxies. Blobby's projected shape appears more circular compared to other Local Volume galaxies. Only about 10 per cent of galaxies in the Catalog of Local Volume Galaxies (Karachentsev & Kaisina 2019) have ellipticities as low or lower than Blobby's measured ellipticity. Outside the Local Volume, Blobby's Sersic index and ellipticity are low compared to blue low surface brightness galaxies (LSBGs) but more typical of red LSBGs (Greco et al. 2018). Simulated red ultra-diffuse galaxies (UDGs) in Benavides et al. (2021) (which includes several examples on backsplash orbits) are more circular than their blue counterparts. All this is to say that although Blobby is a rare example of a backsplash galaxy (or at the very least a relatively isolated and quiescent dwarf), it is structurally similar to low-mass galaxies in other environments. In particular, Blobby is most similar to red dwarf spheroidal populations, which are found in dense environments more often than their blue star-forming counterparts (Weisz et al. 2011a).

## 5 ENVIRONMENT AND INTERPRETATION

We found in Section 2 (Fig. 1) that the M81 group is the closest group in projection to Blobby. In Section 4 (Fig. 6) we showed that the M81 group is the closest in 3D separation and that Blobby's distance and quenched status means it is unlikely to be an unprocessed field galaxy. In this section, we argue that Blobby is likely a backsplash galaxy associated with the system.

The measured SBF distance provides evidence that Blobby is likely within a few $R_{\rm vir}$ of the M81 group and allows us to compare its properties with the few Blobby–M81 backsplash analogues in simulations. For example, Benavides et al. (2021) simulated backsplash UDGs for a range of host masses from galactic haloes (log$M_{200} \sim 12.4$) to cluster haloes (log$M_{200} \sim 14.4$). There are only a few simulated backsplash UDGs in their sample with host haloes roughly similar in mass to M81, and even the smallest UDG stellar mass they consider ($\sim 3 \times 10^7 {\rm M}_\odot$) is larger than the approximate upper limit on Blobby's mass. Blobby also does not qualify as a UDG, since its effective radius is less than 1.5 kpc. However, assuming the simulation results hold broadly given these caveats, the distance we measure from M81 is largely consistent with Benavides et al. (2021) predictions, in which backsplash galaxies around galactic haloes reside at approximately 1–2.5$R_{\rm vir}$ from the host. Similarly, Blobby's measured distance from M81 is consistent with simulated results from Applebaum et al. (2021), who find that backsplash galaxies around MW-like hosts make up over half of the galaxies within 1.5$R_{\rm vir}$ and exist (albeit rarely) out to $\sim 2.5R_{\rm vir}$. Importantly, Blobby's estimated stellar mass puts it near the upper limit of backsplash galaxies in these simulations. More simulations in the mass range where Blobby and M81 reside would help put Blobby's distance in better context.

There are some analogues in Buck et al. (2019) simulations more similar to Blobby's mass, but velocity information, the H I mass or halo mass of Blobby, or a detailed analysis of Blobby's SFH are necessary for comparison. Buck et al. (2019) point out, for instance, that backsplash galaxies (much like satellite galaxies) will have lower gas mass, lower mass-to-light ratios, lower stellar line-of-sight velocity dispersion, and will form their stars earlier than field galaxies. Even without these measurements though, there is evidence to suggest that Blobby has not been forming stars recently and is therefore less likely to be a field galaxy. ArtPop SSP modelling (Section 3.4) indicates that Blobby's stellar population is old ($\gtrsim 10$ Gyr) and the lack of UV emission (Section 4) supports that finding. It is also interesting that Blobby appears to be metal-rich compared to similar galaxies (Fig. 3) since that may be evidence of strangulation from past interactions with a host like M81 (Peng et al. 2015).

Given the current data and the relative rarity of low-mass quenched galaxies like Blobby in the field, we argue that Blobby is more likely a backsplash candidate than a true field galaxy. With this assumption we can make some estimates about Blobby's possible past interactions with the M81 system. If we assume Blobby is moving away from M81 at the escape velocity, $v_{\rm esc} \sim \sqrt{2}v$, where $v$ is the circular velocity at the virial radius of the M81 group, then it would take $\sim 1.5$ Gyr for Blobby to move to its current distance as measured by SBF. If we similarly take into account the uncertainty on Blobby's distance from M81 we estimate Blobby's last interaction was $\sim 1.0$–2.5 Gyr ago, where the minimum time is sharply limited by the minimum possible distance between the galaxies (the projected distance). Smercina et al. (2020) argue that the SFHs of M81's most massive satellites, M82 and NGC 3077, indicate that they began interacting with M81 at the same time $\lesssim 1$ Gyr ago. We examined the possibility that Blobby was initially a satellite of M82 or NGC 3077 since three-body interactions between infalling satellite pairs and a host have the potential to eject the smaller satellite from the system (Sales et al. 2007). If we take Blobby's last interaction with M81 at $\sim 1$ Gyr ago as Smercina et al. (2020) suggests, and assume a constant velocity, we see that Blobby must move at $\sim 1.5v_{\rm esc}$ to travel to its measured distance. This of course is a very order-of-magnitude estimation, so we are not able to rule out the possibility that Blobby was at one time a satellite of M82 or NGC 3077, but given the much smaller simulated radial velocities of similar galaxies in Sales et al. (2007) it is not clear that Blobby was associated with either of these systems. Of course, if Blobby is at the smaller distance suggested by the statistical error then only $\sim 1v_{\rm esc}$ is required, which is more reasonable. There are also hints of an interesting connection between Blobby and other M81 satellites because Smercina et al. (2022) find two low-mass satellites with 90 per cent star formation time-scales very similar to the inferred age of Blobby's stellar population.

Regardless of the precise connection, it seems probable that if the interaction time-scale ($\sim 1.5$ Gyr) is roughly accurate and the SSP modelling is to be trusted ($t_{\rm quench} \gtrsim 10$ Gyr), then Blobby may have been pre-processed before its interaction with M81, meaning it interacted with and was influenced by another galaxy or group of galaxies before falling into the M81 halo. Pre-processing would explain how it was quenched even before first infall (Joshi et al. 2021), something that is not uncommon in the Local Group (Weisz et al. 2015) and happened to about one-third of the backsplash UDGs in Benavides et al. (2021). We can speculate perhaps that the two low-mass M81 satellites in Smercina et al. (2022) with quenching time-scales similar to Blobby could have been involved in the same interactions and fallen into the M81 system together. Confirmation that pre-processing happened because of interactions with M82, NGC 3077, some other galaxy, or not at all, requires further study.

In the future, line-of-sight velocity information would be especially useful to help further constrain the relative motion of Blobby and the M81 group and could help clarify whether the inferred formation path of Blobby is realistic. Other possible avenues for further study include using space-telescope resolved colour–magnitude measurements to characterize Blobby's stellar population (Chiboucas et al. 2013) or measuring Blobby's gas mass to provide further evidence of its interaction history (Buck et al. 2019). In the event that further study confirms the backsplash scenario, Blobby resides in an understudied mass regime and therefore presents an interesting opportunity to further understand dwarf galaxy evolution and quenching mechanisms on time-scales relevant to backsplash populations. If, on the other hand, additional information rejects the backsplash hypothesis then Blobby becomes one of the rare examples of low-mass quenched field galaxies in a mass regime outside the typical influence of the UV background. Either way, Blobby's mass and environment demonstrate that it is an excellent candidate for future study of dwarf galaxy evolution.

## 6 SUMMARY AND CONCLUSION

In this paper, we presented the discovery of dw0910+7326 (Blobby), a new low-mass galaxy (bottom-right panel of Fig. 2) outside the virial radius of the M81 group (Fig. 1). The galaxy was discovered concurrently in Karachentsev & Kaisina (2022). We argue based on the SBF distance and optical and UV properties that it is a good candidate for a backsplash galaxy of the group, meaning it has interacted with the M81 system at some point in the past despite its present isolation. Our main results are as follows.

(i) We use mock galaxies to demonstrate that we can use LBT to measure distances using SBF. The methods used to accomplish this are described in Section 3 and the validation of the method is described in Section 3.4.

(ii) We measure a distance from the Sun of $3.21^{+0.15+0.41}_{-0.15-0.36}$ Mpc compared to $3.67 \pm 0.19$ Mpc for M81 (Section 4). This distance puts Blobby approximately 590 kpc or $1.7 R_{\rm vir}$ from M81 (Fig. 6), consistent with theoretical predictions about the typical distances of backsplash galaxies from their hosts. The projected distance between Blobby and M81 is just over $1.0 R_{\rm vir}$. The maximum distance given uncertainties is $2.7 R_{\rm vir}$. We argue that the SBF-by-eye in Fig. 2 supports the central distance measurement. Table 1 summarizes the measured properties of Blobby.

(iii) Comparisons with ArtPop SSP models suggest that Blobby has a stellar mass of approximately $10^{7.7} M_\odot$ and an old ($\gtrsim 10$ Gyr) stellar population that is fairly metal-rich compared to similar low-mass galaxies (Fig. 3). We argue based on the red ($B - R \sim 1.2$) colour and non-detection in FUV and NUV GALEX images (Section 4) that Blobby is likely quenched. This further reinforces the conclusion that Blobby is a backsplash galaxy because field galaxies with stellar masses $10^{6.5-9} M_\odot$ are nearly all star-forming.

(iv) Rough approximations put Blobby's last interaction with the M81 system of the order of ~1.5 Gyr ago (Section 5). Given the inferred age of Blobby's stellar population ($\gtrsim 10$ Gyr), it is plausible that Blobby was quenched via pre-processing prior to infall into the M81 system since unprocessed field populations in Blobby's mass range are primarily star-forming. The timeline roughly matches up with the $\lesssim 1$ Gyr interaction timeline of M81 and its two most massive neighbours (M82 and NGC 3077) predicted by Smercina et al. (2020). It is possible that Blobby was pre-processed by one of those systems, although more study into this hypothesis is required. If Blobby was quenched by M81 directly its stellar population would likely be younger, which would require a higher metallicity given its colour and make Blobby even more of an outlier on the MZR (Fig. 3). The age of Blobby's stellar population also coincides with the 90 per cent star formation time-scales for two of M81's other low-mass satellites (Smercina et al. 2022).

This is the first result from our campaign to measure distances to candidate Local Volume dwarf galaxies with the LBT. One of the primary strengths of the SBF method is the ability to measure accurate distances given photometric data alone, in this case confirming that Blobby is within a few $R_{\rm vir}$ of the M81 system. However, obtaining velocity information for Blobby would go a long way to untangle the remaining uncertainty with regard to its interaction with M81.

## ACKNOWLEDGEMENTS

This work depended on the feedback and support of many, including Dr Richard Pogge regarding image processing, Dr Morgan Schmitz and Dr Lee Kelvin on star subtraction, Dr Scott Carlsten on SBF, Dr Paul Martini on uncertainty analysis and dwarf galaxy comparisons, and Dr Adam Smercina for helpful comments regarding the possible connection with some of M81's other satellites. Thank you to Dr Pieter van Dokkum for help developing our observing strategy. Thank you also to Dr Shany Danieli for her significant contributions to ArtPop which greatly improved this work. The first author would also like to extend a heartfelt thank you to the technical support staff at the Ohio Supercomputer Center (Ohio Supercomputer Center 1987) for their patience helping troubleshoot and improve the pipeline used for the computations in this paper.

The LBT is an international collaboration among institutions in the United States, Italy, and Germany. LBT Corporation partners are: The University of Arizona on behalf of the Arizona Board of Regents; Istituto Nazionale di Astrofisica, Italy; LBT Beteiligungsgesellschaft, Germany, representing the Max-Planck Society, The Leibniz Institute for Astrophysics Potsdam, and Heidelberg University; The Ohio State University, representing OSU, University of Notre Dame, University of Minnesota, and University of Virginia.

The Pan-STARRS1 Surveys (PS1) and the PS1 public science archive have been made possible through contributions by the Institute for Astronomy, the University of Hawaii, the Pan-STARRS Project Office, the Max-Planck Society and its participating institutes, the Max Planck Institute for Astronomy, Heidelberg and the Max Planck Institute for Extraterrestrial Physics, Garching, The Johns Hopkins University, Durham University, the University of Edinburgh, the Queen's University Belfast, the Harvard-Smithsonian Center for Astrophysics, the Las Cumbres Observatory Global Telescope Network Incorporated, the National Central University of Taiwan, the Space Telescope Science Institute, the National Aeronautics and Space Administration under Grant No. NNX08AR22G issued through the Planetary Science Division of the NASA Science Mission Directorate, the National Science Foundation Grant No. AST-1238877, the University of Maryland, Eotvos Lorand University (ELTE), the Los Alamos National Laboratory, and the Gordon and Betty Moore Foundation.

This research has used the NASA/IPAC Extragalactic Database (NED), which is funded by the National Aeronautics and Space Administration and operated by the California Institute of Technology. This research has also used the NASA/IPAC Infrared Science Archive, which is funded by the National Aeronautics and Space Administration and operated by the California Institute of Technology.

Some of the data presented in this paper were obtained from the Mikulski Archive for Space Telescopes (MAST). STScI is operated by the Association of Universities for Research in Astronomy, Inc., under NASA contract NAS5-26555. Support for MAST for non-HST data is provided by the NASA Office of Space Science via grant NNX13AC07G and by other grants and contracts.

This research used observations made with the NASA Galaxy Evolution Explorer. GALEX is operated for NASA by the California Institute of Technology under NASA contract NAS5-98034.

The software used in this work includes ARTPOP (Greco & Danieli 2021), IMFIT (Erwin 2015), ASTROPY (Astropy Collaboration 2013, 2018), NUMPY (Harris et al. 2020), MATPLOTLIB (Hunter 2007), SOURCE EXTRACTOR (Bertin & Arnouts 1996), SEP (Barbary 2016), SCIPY (Virtanen et al. 2020), PHOTUTILS (Bradley et al. 2020), CCDPROC (Craig et al. 2017), and ASTROMETRY.NET (Lang et al. 2010). This research has used NASA's Astrophysics Data System.

KJC, AHGP, and ABD are supported by NSF Grant No. AST-1615838. KJC and AHGP are additionally supported by NSF Grant No. AST-2008110. JPG is supported by an NSF Astronomy and Astrophysics Postdoctoral Fellowship under award AST-1801921.

## DATA AVAILABILITY

The data underlying this article will be shared on reasonable request to the corresponding author.

## REFERENCES

Akins H. B., Christensen C. R., Brooks A. M., Munshi F., Applebaum E., Engelhardt A., Chamberland L., 2021, ApJ, 909, 139
Applebaum E., Brooks A. M., Christensen C. R., Munshi F., Quinn T. R., Shen S., Tremmel M., 2021, ApJ, 906, 96
Astropy Collaboration, 2013, A&A, 558, A33
Astropy Collaboration, 2018, AJ, 156, 123
Bakels L., Ludlow A. D., Power C., 2021, MNRAS, 501, 5948
Barbary K., 2016, J. Open Source Softw., 1, 58
Benavides J. A. et al., 2021, Nat. Astron., 5, 1255
Benavides J. A., Sales L. V., Abadi M., Marinacci F., Vogelsberger M., Hernquist L., 2022, preprint (arXiv:2209.07539)
Bertelli G., Bressan A., Chiosi C., Fagotto F., Nasi E., 1994, A&AS, 106, 275
Bertin E., Arnouts S., 1996, A&AS, 117, 393
Blakeslee J. P., 2012, Ap&SS, 341, 179
Blakeslee J. P., Ajhar E. A., Tonry J. L., 1999, Post-Hipparcos Cosmic Candles. Springer, Netherlands, p. 181
Blaña M., Burkert A., Fellhauer M., Schartmann M., Alig C., 2020, MNRAS, 497, 3601
Bluck A. F. L. et al., 2020, MNRAS, 499, 230
Borrow J., Vogelsberger M., O'Neil S., McDonald M. A., Smith A., 2023, MNRAS, 520, 649
Boselli A., Gavazzi G., 2006, PASP, 118, 517
Bradley L. et al., 2020, astropy/photutils: 1.0.0, https://doi.org/10.5281/zenodo.4044744
Buck T., Macciò A. V., Dutton A. A., Obreja A., Frings J., 2019, MNRAS, 483, 1314
Cantiello M. et al., 2018, ApJ, 856, 126
Carlsten S. G., Beaton R. L., Greco J. P., Greene J. E., 2019, ApJ, 879, 13
Carlsten S. G., Greene J. E., Greco J. P., Beaton R. L., Kado-Fong E., 2021a, preprint (arXiv:2105.03435)
Carlsten S. G., Greene J. E., Greco J. P., Beaton R. L., Kado-Fong E., 2021c, ApJ, 922, 267
Carlsten S. G., Greene J. E., Peter A. H., Beaton R. L., Greco J. P., 2021b, ApJ, 908, 109
Chambers K. C. et al., 2016, preprint (arXiv:1612.05560)
Chiboucas K., Jacobs B. A., Tully R. B., Karachentsev I. D., 2013, AJ, 146, 126
Choi J., Dotter A., Conroy C., Cantiello M., Paxton B., Johnson B. D., 2016, ApJ, 823, 102
Christensen C. R., Davé R., Governato F., Pontzen A., Brooks A., Munshi F., Quinn T., Wadsley J., 2016, ApJ, 824, 57
Cohen Y. et al., 2018, ApJ, 868, 96
Cortese L., Catinella B., Smith R., 2021, Publ. Astron. Soc. Aust., 38, e035
Craig M. et al., 2017, astropy/ccdproc: v1.3.0.post1, https://doi.org/10.5281/zenodo.1069648
Croton D. J. et al., 2006, MNRAS, 365, 11
Danieli S., van Dokkum P., Conroy C., 2018, ApJ, 856, 69
Davis A. B. et al., 2021, MNRAS, 500, 3854
Dey A. et al., 2019, AJ, 157, 168
Diemer B., 2021, ApJ, 909, 112
Dotter A., 2016, ApJS, 222, 8
Du W., McGaugh S. S., 2020, AJ, 160, 122
Erwin P., 2015, ApJ, 799, 226
Fillingham S. P., Cooper M. C., Boylan-Kolchin M., Bullock J. S., Garrison-Kimmel S., Wheeler C., 2018, MNRAS, 477, 4491
Fillingham S. P., Cooper M. C., Wheeler C., Garrison-Kimmel S., Boylan-Kolchin M., Bullock J. S., 2015, MNRAS, 454, 2039
Fitts A. et al., 2017, MNRAS, 471, 3547
Flewelling H. E. et al., 2020, ApJS, 251, 7
Forbes D. A., Read J. I., Gieles M., Collins M. L. M., 2018, MNRAS, 481, 5592
Gabor J. M., Davé R., Finlator K., Oppenheimer B. D., 2010, MNRAS, 407, 749
García-Benito R., Delgado R. G., Pérez E., Fernandes R. C., Sánchez S., de Amorim A., 2019, A&A, 621, A120
Garling C. T., Peter A. H. G., Kochanek C. S., Sand D. J., Crnojević D., 2020, MNRAS, 492, 1713
Garrison-Kimmel S., Boylan-Kolchin M., Bullock J. S., Lee K., 2014, MNRAS, 438, 2578
Geha M., Blanton M. R., Yan R., Tinker J. L., 2012, ApJ, 757, 85
Gill S. P. D., Knebe A., Gibson B. K., 2005, MNRAS, 356, 1327
Gnedin N. Y., Kravtsov A. V., 2006, ApJ, 645, 1054
Graham A. W., Driver S. P., 2005, Publ. Astron. Soc. Aust., 22, 118
Grebel E. K., Gallagher III J. S., Harbeck D., 2003, AJ, 125, 1926
Greco J. P. et al., 2018, ApJ, 857, 104
Greco J. P., Danieli S., 2021, ApJ, 941, 26
Greco J. P., van Dokkum P., Danieli S., Carlsten S. G., Conroy C., 2021, ApJ, 908, 24
Greene J. E., Danieli S., Carlsten S., Beaton R., Jiang F., 2022, preprint (arXiv:2210.14237)
Gunn J. E., Gott III J. R., 1972, ApJ, 176, 1
Guo H., Jones M. G., Wang J., Lin L., 2021, ApJ, 918, 53
Haggar R., Gray M. E., Pearce F. R., Knebe A., Cui W., Mostoghiu R., Yepes G., 2020, MNRAS, 492, 6074
Harris C. R. et al., 2020, Nature, 585, 357
Hill J. M., 2010, Appl. Opt., 49, D115
Hopkins P. F., Quataert E., Murray N., 2012, MNRAS, 421, 3522
Hunter J. D., 2007, Comput. Sci. Eng., 9, 90
IRSA, 2022, Galactic Dust Reddening and Extinction, IPAC
Jahn E. D., Sales L. V., Wetzel A., Samuel J., El-Badry K., Boylan-Kolchin M., Bullock J. S., 2022, MNRAS, 513, 2673
Jensen J. B., Tonry J. L., Luppino G. A., 1998, ApJ, 505, 111
Jeon M., Besla G., Bromm V., 2017, ApJ, 848, 85
Jerjen H., Binggeli B., Barazza F. D., 2004, AJ, 127, 771
Jerjen H., Freeman K. C., Binggeli B., 2000, AJ, 119, 166
Jester S. et al., 2005, AJ, 130, 873
Joshi G. D., Pillepich A., Nelson D., Zinger E., Marinacci F., Springel V., Vogelsberger M., Hernquist L., 2021, MNRAS, 508, 1652
Karachentsev I. D. et al., 2001, A&A, 379, 407
Karachentsev I. D., Kaisina E., 2019, Astrophys. Bull., 74, 111
Karachentsev I. D., Kniazev A. Y., Sharina M. E., 2015, Astron. Nachr., 336, 707
Karachentsev I. D., Kudrya Y. N., 2014, AJ, 148, 50
Karachentsev I., Kaisina E., 2022, Astrophys. Bull., 77, 372
Kirby E. N., Cohen J. G., Guhathakurta P., Cheng L., Bullock J. S., Gallazzi A., 2013, ApJ, 779, 102
Kroupa P., 2001, MNRAS, 322, 231
Lang D., Hogg D. W., Mierle K., Blanton M., Roweis S., 2010, AJ, 139, 1782
Larson R., Tinsley B., Caldwell C. N., 1980, ApJ, 237, 692
Lavery R. J., Mighell K. J., 1992, AJ, 103, 81
Leroy A. K. et al., 2019, ApJS, 244, 24
Li P., Wang H., Mo H. J., Wang E., Hong H., 2020, ApJ, 902, 75
Lotz M., Remus R.-S., Dolag K., Biviano A., Burkert A., 2019, MNRAS, 488, 5370
Lu Y., Benson A., Wetzel A., Mao Y.-Y., Tonnesen S., Peter A. H. G., Boylan-Kolchin M., Wechsler R. H., 2017, ApJ, 846, 66
Magnier E. A. et al., 2020, ApJS, 251, 6
Makarov D., Makarova L., Sharina M., Uklein R., Tikhonov A., Guhathakurta P., Kirby E., Terekhova N., 2012, MNRAS, 425, 709
Makarova L. N., Makarov D. I., Karachentsev I. D., Tully R. B., Rizzi L., 2017, MNRAS, 464, 2281
Mao Y.-Y., Geha M., Wechsler R. H., Weiner B., Tollerud E. J., Nadler E. O., Kallivayalil N., 2021, ApJ, 907, 85
Martínez-Delgado D., Alonso-García J., Aparicio A., Gómez-Flechoso M. A., 2001, ApJ, 549, L63
Mayer L., Kazantzidis S., Mastropietro C., Wadsley J., 2007, Nature, 445, 738
McConnachie A. W. et al., 2008, ApJ, 688, 1009

McConnachie A. W., 2012, AJ, 144, 4
McGaugh S. S., Schombert J. M., 2014, AJ, 148, 77
Mei S. et al., 2005, ApJ, 625, 121
Miller B. W., Lotz J. M., 2007, ApJ, 670, 1074
More S., Diemer B., Kravtsov A. V., 2015, ApJ, 810, 36
Moster B. P., Naab T., White S. D. M., 2013, MNRAS, 428, 3121
Munshi F. et al., 2013, ApJ, 766, 56
Ohio Supercomputer Center, 1987, Ohio Supercomputer Center , http://osc.edu/ark:/19495/f5s1ph73
Oke J. B., Gunn J. E., 1983, ApJ, 266, 713
Oman K. A., Bahé Y. M., Healy J., Hess K. M., Hudson M. J., Verheijen M. A. W., 2021, MNRAS, 501, 5073
Peng Y., Maiolino R., Cochrane R., 2015, Nature, 521, 192
Pimbblet K. A., 2011, MNRAS, 411, 2637
Polzin A., van Dokkum P., Danieli S., Greco J. P., Romanowsky A. J., 2021, ApJ, 914, L23
Ragazzoni R. et al., 2006, in Stepp L. M., ed., Ground-based and Airborne Telescopes, vol. 6267. SPIE, Orlando, Florida, United States, p. 348
Rey M. P., Pontzen A., Agertz O., Orkney M. D. A., Read J. I., Rosdahl J., 2020, MNRAS, 497, 1508
Rey M. P., Pontzen A., Agertz O., Orkney M. D. A., Read J. I., Saintonge A., Kim S. Y., Das P., 2022, MNRAS, 511, 5672
Ricotti M., Gnedin N. Y., 2005, Proc. Int. Astron. Union, 1, 77
Rines K., Geller M. J., Kurtz M. J., Diaferio A., 2005, AJ, 130, 1482
Sales L. V., Navarro J. F., Abadi M. G., Steinmetz M., 2007, MNRAS, 379, 1475
Samuel J., Wetzel A., Santistevan I., Tollerud E., Moreno J., Boylan-Kolchin M., Bailin J., Pardasani B., 2022, MNRAS, 514, 5276
Sand D. J. et al., 2022, ApJ, 935, L17
Sand D. J., Seth A., Olszewski E. W., Willman B., Zaritsky D., Kallivayalil N., 2010, ApJ, 718, 530
Santos-Santos I. M., Navarro J. F., McConnachie A., 2023, MNRAS, 520, 55
Schlafly E. F., Finkbeiner D. P., 2011, ApJ, 737, 103
Sharina M. E. et al., 2008, MNRAS, 384, 1544
Simpson C. M., Grand R. J. J., Gómez F. A., Marinacci F., Pakmor R., Springel V., Campbell D. J. R., Frenk C. S., 2018, MNRAS, 478, 548
Smercina A. et al., 2020, ApJ, 905, 60
Smercina A., Bell E. F., Samuel J., D'Souza R., 2022, ApJ, 930, 69
Speziali R. et al., 2008, in McLean I. S., Casali M. M., eds, Ground-based and Airborne Instrumentation for Astronomy II, 7014. SPIE, Marseille, France, p. 1663
Team COSMOS, 2020, GALEX/COSMOS Prior-based Photometry Catalog. IPAC
Teyssier M., Johnston K. V., Kuhlen M., 2012, MNRAS, 426, 1808
Tonry J. L., Ajhar E. A., Luppino G. A., 1990, AJ, 100, 1416
Tonry J., Schneider D. P., 1988, AJ, 96, 807
Trussler J., Maiolino R., Maraston C., Peng Y., Thomas D., Goddard D., Lian J., 2020, MNRAS, 491, 5406
Trussler J., Maiolino R., Maraston C., Peng Y., Thomas D., Goddard D., Lian J., 2021, MNRAS, 500, 4469
van Dokkum P., Danieli S., Cohen Y., Romanowsky A. J., Conroy C., 2018, ApJ, 864, L18
Virtanen P. et al., 2020, Nat. Methods, 17, 261
Waters C. Z. et al., 2020, ApJS, 251, 4
Weisz D. R. et al., 2011a, ApJ, 739, 5
Weisz D. R. et al., 2011b, ApJ, 743, 8
Weisz D. R., Dolphin A. E., Skillman E. D., Holtzman J., Gilbert K. M., Dalcanton J. J., Williams B. F., 2015, ApJ, 804, 136
Wenger M. et al., 2000, A&AS, 143, 9
Wetzel A. R., Tinker J. L., Conroy C., Bosch F. C. v. d., 2014, MNRAS, 439, 2687
Wetzel A. R., Tinker J. L., Conroy C., van den Bosch F. C., 2013, MNRAS, 432, 336
Wheeler C., Phillips J. I., Cooper M. C., Boylan-Kolchin M., Bullock J. S., 2014, MNRAS, 442, 1396
Whiting A. B., Hau G. K. T., Irwin M., 1999, AJ, 118, 2767
Willmer C. N., 2018, ApJS, 236, 47
Worthey G., 1994, ApJS, 95, 107
Wright R. J., de Lagos C. P., Power C., Stevens A. R. H., Cortese L., Poulton R. J. J., 2022, MNRAS, 516, 2891

## APPENDIX A: THE CHALLENGE OF BD+73 447

As we discussed in Section 2, there is an unfortunately bright star ($m_R \sim 9$ and $m_B \sim 11$) located approximately 2 arcmin from the centre of dw0910+7326. Figs 2 and 4 show images where the star has been removed using six stacked Gaussian kernels in each of the 10 exposures. Fig. A1 shows the same image where the star was not subtracted. There is significant contamination of diffuse light from the star overlapping the galaxy that is not present in the star subtracted images of Figs 2 and 4.

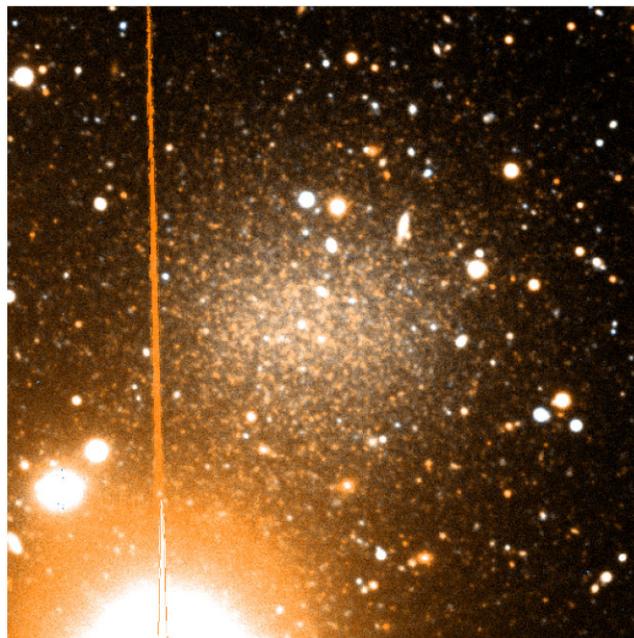

**Figure A1.** RGB image of dw0910+7326 before star subtraction using the same image parameters as Fig. 2. The RGB image was created using *R*- and *B*-band images where the average of the two bands was used for the green channel. The star (BD+73 447) is approximately ninth magnitude in *R*- band and 11th magnitude in *B*- band (Wenger et al. 2000). Section 2 describes the star subtraction utilized in this work.

This paper has been typeset from a T$_{\rm E}$X/L$^{\rm A}$T$_{\rm E}$X file prepared by the author.